\documentclass[aps,prl,twocolumn,superscriptaddress,showpacs]{revtex4}
\usepackage{graphicx}
\usepackage{latexsym}
\usepackage{amssymb}
\usepackage{amsmath}
\usepackage{amsfonts}
\usepackage{multirow}
\usepackage{color}
\usepackage{comment}

\newcommand{\U}{\mathrm{U}}
\newcommand{\Sp}{\mathrm{Sp}}
\newcommand{\Gr}{\mathrm{Gr}}

\newcommand{\eqnref}[1]{Eq.\,\eqref{#1}}

\newcommand{\beq}{\begin{equation}}
\newcommand{\eeq}{\end{equation}}
\newcommand{\beqn}{\begin{eqnarray}}
\newcommand{\eeqn}{\end{eqnarray}}

\begin{document}

\title{Symmetry Protected Topological Hopf Insulator and its Generalizations}

%\author{Cenke's group}

\author{Chunxiao Liu}

\author{Farzan Vafa}

\author{Cenke Xu}

\affiliation{Department of physics, University of California,
Santa Barbara, CA 93106, USA}

\begin{abstract}

We study a class of $3d$ and $4d$ topological insulators whose
topological nature is characterized by the Hopf map and its
generalizations. We identify the symmetry $\mathcal{C}^\prime$, a
generalized particle-hole symmetry that gives the Hopf insulator a
$\mathbb{Z}_2$ classification. The $4d$ analogue of the Hopf
insulator with symmetry $\mathcal{C}^\prime$ has the same
$\mathbb{Z}_2$ classification. The minimal models for the $3d$ and
$4d$ Hopf insulator can be heuristically viewed as
``Chern-insulator$ \rtimes S^1$" and ``Chern-insulator$\rtimes
T^2$" respectively. We also discuss the relation between the Hopf
insulator and the Weyl semimetals, which points the direction for
its possible experimental realization.

\end{abstract}


\maketitle

%{\it --- Introduction}

The 10-fold way classification has provided us the prototypes of
topological insulators and topological
superconductors~\cite{kitaevclass,ludwigclass1,ludwigclass2}. The
usual wisdom is that, even the topological insulators with
symmetries beyond the 10-fold way classification can also be
understood as these prototypes enriched with other symmetries.
Depending on the dimension and symmetry, the boundary states of
all these prototypes should be a gapless Dirac fermion, or Weyl
fermion, or Majorana fermion. One important open question is, can
these prototypes represent all possible topological insulators, or
can we still find exceptions that are different from states in the
10-fold way classification? In this paper we present a class of
such examples, which we generally call Hopf insulator.

The Hopf insulator was studied in
Ref.~\onlinecite{Moore2008,Deng2013}, but the key symmetry that
protects the Hopf insulator was not identified. Without a proper
symmetry, the Hopf insulator is actually trivial, which we will
explain later in this paper. We demonstrate that the topological
nature of the $3d$ Hopf insulator state, previously described only
by the homotopy group $\pi_3[S^2] = Z$ in the simplest two band
model~\cite{Moore2008,Deng2013}, is more generally given by $\pi_3
[\mathrm{Sp}(2N)/ \mathrm{U}(N)] = Z_2$ in a multi-band
model~\cite{convention}, which is the key to stabilizing this
state under physical conditions. As long as the system has the
translation and a $\mathcal{C}^\prime$ particle-hole symmetry (to
be defined later), this Hopf insulator has a $\mathbb{Z}_2$
classification. The same symmetry $\mathcal{C}^\prime$ also gives
us a $4d$ topological insulator based on the homotopy group
$\pi_4[S^2] = Z_2$ and $\pi_4[\Sp(2N)/\U(N)] = Z_2$.

\begin{figure}[tbp]
\begin{center}
\includegraphics[width=130pt]{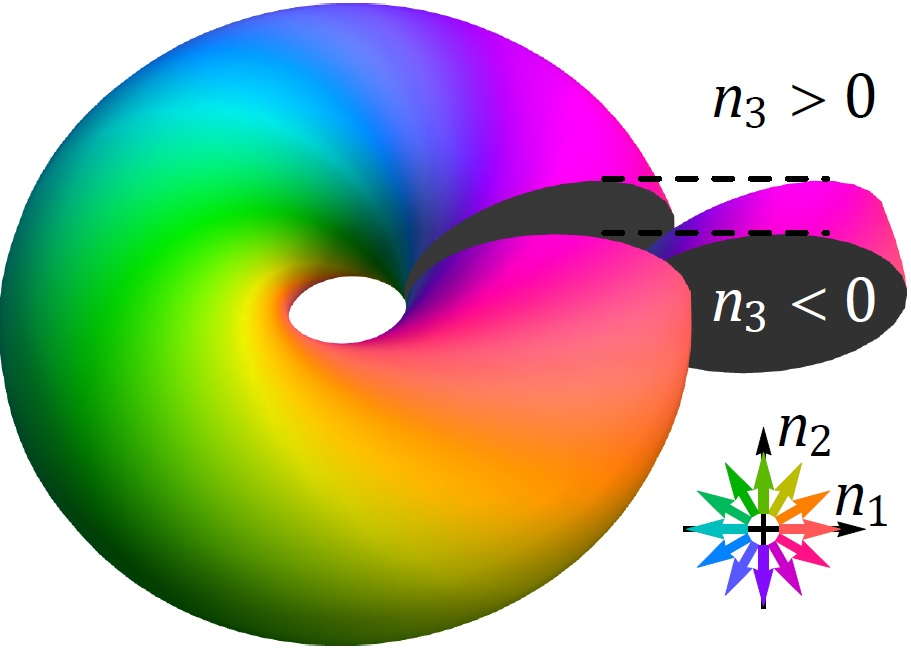}
\caption{An illustration of the Hopf map in the momentum space.
The domain wall $n_3 = 0$ forms a solid torus, and $(n_1, n_2)$
winds around both directions of the torus. Thus intuitively the
Hopf insulator can be viewed as ``Chern-insulator$\rtimes S^1$".
The symbol $\rtimes$ represents the ``winding of the
Chern-insulator" along $S^1$.} \label{hopf}
\end{center}
\end{figure}

The minimal model of the Hopf insulator was constructed in
Ref.~\onlinecite{Moore2008,Deng2013}, but one has to study it with
caution. The Brillouin zone of a $3d$ lattice model is a
three-torus $T^3$. A topologically nontrivial mapping from $T^3$
to $S^2$ is equivalent to the mapping from $S^3$ to $S^2$ ($i.e.$
the so called Hopf map), as long as the target manifold $S^2$ has
zero winding on any two-torus submanifold of the Brillouin zone.
This Hopf map can be induced from the standard mapping from $T^3$
to $S^3$ with winding number $1$. Thus the simplest Hopf insulator
Hamiltonian is a two-band model in the $3d$ Brillouin
zone~\cite{Moore2008,Deng2013}: \beqn H(\bold k) = \vec{n}(\bold
k)\cdot \vec{\sigma}. \label{2band}\eeqn The three component
vector field $\vec{n}(\bold k)$ is a mapping from the Brillouin
zone to the target space $S^2$, and we choose it to have Hopf
number $+1$. The schematic configuration of $\vec{n}$ in the
momentum space is depicted in Fig.~\ref{hopf}.

\begin{figure}[!thb]
\centering
\includegraphics[width=170pt]{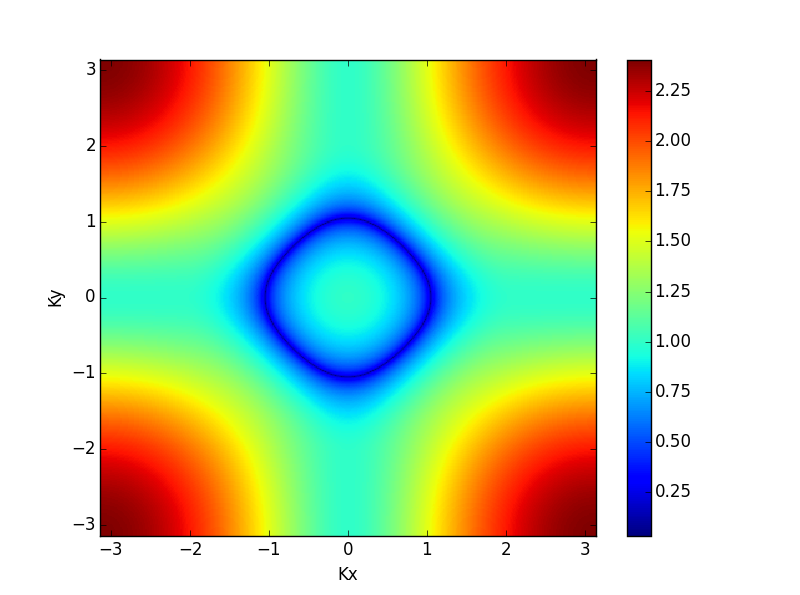}
\caption{The Fermi ring at the boundary of the minimal two-band
model of the Hopf insulator for $m=2$ in
Eq.~\ref{N}.}\label{fermiring}
\end{figure}

We can introduce the standard CP$^1$ field $z(\bold k) =
(z_1(\bold k), \ z_2(\bold k))^t$ for the three component vector
$\vec{n}(\bold k)$: \beqn \vec{n}(\bold k) = z^\dagger (\bold k)
\vec{\sigma} z(\bold k), \eeqn the spectrum of the Hamiltonian is
\beqn E(\bold k) = \pm |\vec{n}(\bold k)| = \pm |z(\bold k)|^2 =
\pm |\vec{N}(\bold k)|^2, \eeqn where $\vec{N}(\bold k) $ is a
four component vector defined as $z_1(\bold k) = N_1(\bold k) + i
N_2(\bold k)$, $z_2 = N_3(\bold k) + i N_4(\bold k)$. Thus as long
as the length of $\vec{N}$ never vanishes in the Brillouin zone,
this system is an insulator with a band gap between the two bands.
The configuration of $\vec{n}(\bold k)$ with Hopf number $+1$,
corresponds to the configuration of $\vec{N}(\bold k)$ with the
winding number $+1$: \beqn \mathrm{Hopf \ number} =
\frac{1}{\Omega_3} \int d^3k \ \hat{N}^a \partial_{k_x} \hat{N}^b
\partial_{k_y} \hat{N}^c
\partial_{k_z} \hat{N}^d, \label{hopfn}\eeqn where $\hat{N}(\bold k) =
\vec{N}(\bold k)/|\vec{N}(\bold k)|$, and $\Omega_3$ is the volume
of $S^3$.

As an example of Hopf insulator, we can choose the following
configuration of $\vec{N}(\bold k)$: \beqn && N_1(\bold k) =
\sin(k_x), \ \ N_1(\bold k) = \sin(k_y), \ \ N_3(\bold k) =
\sin(k_z), \cr\cr && N_4(\bold k) = m - \cos(k_x) - \cos(k_y) -
\cos(k_z), \label{N} \eeqn the Hopf number defined in
Eq.~\ref{hopfn} equals $+1$ when $1 < m < 3$. This model is
essentially the same model constructed in
Ref.~\onlinecite{Moore2008,Deng2013}. This two-band model alone
appears to be a nontrivial topological insulator with stable edge
states. The existence of edge states was demonstrated numerically
in Ref.~\onlinecite{Moore2008,Deng2013}, and it was shown that the
edge state of this model is a ``Fermi ring".

The boundary Fermi ring can be heuristically understood as
follows: We can parameterize the $3d$ momentum space as $(k_r,
\theta, k_z)$, where $k_r = \sqrt{k_x^2 + k_y^2}$, and $\tan
\theta = k_y/k_x$. Fig.~\ref{hopf} shows that for every half plane
$(k_r, k_z)$ with fixed $\theta$, $\vec{n}(\bold k)$ has a
configuration with Skyrmion number $+1$, thus for every $\theta$
with $ 0< \theta < 2\pi$, there is a gapless edge state along the
radial $k_r$ direction at the $(0,0,1)$ boundary. These edge
states together will form a Fermi ring on the $(0,0,1)$ boundary.
This observation is confirmed by our direct numerical calculation,
see Fig. \ref{fermiring}. In fact, the Hopf mapping corresponds to
the configuration of $\vec{n}$ in the $3d$ Brillouin zone such
that the domain wall between $n_3 > 0$ and $n_3 < 0$ forms a
torus, and the two component vector $(n_1, n_2)$ has a nontrivial
winding around both directions of the torus. So in this sense, we
can call the $3d$ Hopf insulator as ``Chern-insulator $\rtimes
S^1$", where $\rtimes$ represents the winding of $(n_1, n_2)$
along $S^1$, so the Hopf insulator is not a simple direct product
between the Chern-insulator and $S^1$.

However, this simple two-band Hopf insulator, without any
symmetry, is unstable against mixing with other bands. A generic
band insulator consists of $m$ empty bands and $n$ filled bands,
and in $\bold k$ space can be described by an $(m + n) \times (m +
n)$ Hermitian matrix $H(\bold k)$. For each value of $\bold k$,
$H(\bold k)$ has $m$ positive eigenvalues, corresponding to the
$m$ empty bands, and $n$ negative eigenvalues, corresponding to
the $n$ filled bands. Without closing the gap, the $m$ positive
eigenvalues ($n$ negative eigenvalues) can all be continuously
deformed to $+1$ ($-1$). Therefore, $H(\bold k)$ takes the form $
H(\bold k) = U(\bold k) I_{m,n} U^\dag (\bold k)$, where $I_{m,n}$
is an $(m + n) \times (m + n)$ diagonal matrix with $m$ $+1$'s and
$n$ $-1$'s on the diagonal, and $U(\bold k) \in \U(m + n)$. When
there is no symmetry other than the $U(1)$ charge conservation and
momentum conservation, the configuration space of the Hamiltonian
is topologically equivalent to the complex Grassmannian manifold
$\Gr(n,m+n) = \frac{\U(m + n)}{\U(m) \times
\U(n)}$~\cite{Moore2008}. The band insulator is a map from the
Brillouin zone $T^d$ to $\Gr(n,m+n)$, which is related to the
torus homotopy group $\tau_d[\Gr(n,m+n)]$~\cite{Fox1948}. Since
classes of mappings $T^d \to \Gr(n,m+n)$ are induced by classes of
mappings $S^d \to \Gr(n,m+n)$, we shall focus on the latter.

In general $\pi_3[\Gr(n, n+m)] = 0$, as long as $n$ and $m$ do not
both equal $1$. This observation implies that once the two-band model
described above starts mixing with other bands, there will be no
nontrivial topological insulator. But in the following we will
prove that with a special symmetry $\mathcal{C}^\prime$, this
system always has an even number of bands, and its Hamiltonian
belongs to the manifold $\mathcal{M} = \Sp(2N)/\U(N)$, and because
$\pi_3[\Sp(2N)/\U(N)] = Z_2$ for $N
> 1$, the Hopf insulator has $\mathbb{Z}_2$ classification with
symmetry $\mathcal{C}^\prime$.

For a $2N$-band system, the symmetry $\mathcal C'$ acts on fermion
operators as $\mathcal{C}' f_{k} \mathcal{C}'^{-1}
 = J f_k^\dagger$. $J$ is a $2N \times 2N $ matrix which we choose to be \beqn J =
\begin{pmatrix} 0 & I_{N\times N}\\ -I_{N\times N} & 0
\end{pmatrix}. \eeqn Thus $\mathcal C'$ is a generalized particle-hole transformation,
and it can be viewed as the product between a particle-hole
transformation $\mathcal{C}$ (with $\mathcal{C}^2 = - 1$) and the
spatial inversion $\mathcal{I}$. The symmetry $\mathcal C'$
implies that for all $\bold k$, the Hamiltonian $H(\bold k)$ must
satisfy \beqn \label{C'symmetry} J^{-1} H(\bold k) J = -H(\bold
k)^\ast. \eeqn \eqnref{C'symmetry} implies that for each $\bold
k$, $H(\bold k)$ is in the Lie algebra of $\Sp(2N)$, and there is
always an even number of bands. When diagonalized, $H(\bold k)$
takes the form
\begin{align}
&H(\bold k) = \nonumber \\
&\text{diag}(\lambda_1(\bold k),-\lambda_1(\bold k),
\lambda_2(\bold k), -\lambda_2(\bold k), \ldots, \lambda_N(\bold
k), -\lambda_N(\bold k)).
\end{align}
We can continuously deform all of the positive eigenvalues to
$+1$, and all of the negative eigenvalues to $-1$. Therefore, the
deformed $H(\bold k)$ takes the form
\beq H(\bold k) = U(\bold k) I_{N,N} U^\dag (\bold k), \eeq
where $U(\bold k) \in \Sp(2N)$. We now show that the entire
configuration space of the Hamiltonian is $\Sp(2N)/\U(N)$. A
generic element that does not move $I_{N,N}$ is $g =
\text{diag}(U_1(\bold k), U_2(\bold k))$, where $U_1(\bold k),
U_2(\bold k) \in \U(N)$. However, in order for $g$ to be in
$\Sp(2N)$, it has to satisfy
\beq \begin{pmatrix} U_1 & \\ & U_2 \end{pmatrix} J
\begin{pmatrix} U_1 & \\ & U_2 \end{pmatrix}^T = J. \eeq
Imposing this condition tells us that $U_1 U_2^T = 1$, which
implies that the configuration space of the Hamiltonian is
$\Sp(2N)/\U(N)$. From the mathematics literature,
\begin{subequations}
\begin{eqnarray}
\pi_3 [\Sp(2N)/\U(N)] &=& \begin{cases} Z & N = 1 \\ \\ Z_2 & N
\ge 2\end{cases} \\ \\ \pi_4[\Sp(2N)/\U(N)] &=& \quad Z_2
\end{eqnarray}
\end{subequations}

The conclusion that the classification for the multi-band system
with the $\mathcal{C}^\prime$ symmetry is no larger than $\mathbb
Z_2$ can be understood as follows. Let us take $N=2$. The
nontrivial mapping from $S^3$ to $\Sp(4)/\U(2)$ is induced by
the mapping from $S^3$ to $\Sp(4)$, characterized by the winding
number $n_w = \frac{1}{24\pi^2} \int d^3 k (U^{-1}dU)^3$, where $U
\in \Sp(4)$. When $n_w = 2$, the induced Hamiltonian $ H(\bold k)
= U(\bold k) I_{N,N} U^\dag (\bold k)$ can be continuously
deformed into two copies of two-band Hopf insulators each with the
Hamiltonian Eq.\eqref{2band}, whose three-component vector
$\vec{n}(\bold k)_1 = - \vec{n}(\bold k)_2$ (one can check that
$\vec{n}(\bold k)$ and $- \vec{n}(\bold k)$ give the same Hopf
number). Then it is obvious that this state should not have any
edge state at the $(0,0,1)$ boundary because each $(k_r, k_z)$
half-plane for a fixed $\theta$ now has zero Skyrmion number.

The classification for $\mathcal C^\prime$ can be understood in
another way. Under ordinary $\mathcal C$ transformation, the
position variables are invariant, but all of the momenta variables
pick up a sign. However, under the $\mathcal C^\prime$
transformation, the momenta do not change, but all of the position
variables pick up a sign. Therefore, for all intents and purposes,
we can replace $\mathcal C^\prime$ with $\mathcal C$ as long as we
reverse the roles of the position and momenta variables.
Therefore, following Ref.~\onlinecite{TeoKane}, the classification
of a $d$-dimensional TI with $\mathcal C^\prime$ is the same as
the classification of a $ \delta = 0-d\equiv 8-d (\mathrm{mod} \
8)$-dimensional TI with $\mathcal C$, which is simply Class C. As
expected, the classifications match for all $d$.

We note that because the general multi-band model of the Hopf
insulator requires the $\mathcal C^\prime$ symmetry which involves
spatial inversion, the boundary of the system necessarily breaks
$\mathcal{C}^\prime$ and hence the system does not have protected
edge states. However, the classification of the Hopf insulator is
still well-defined in the bulk, like all the topological
insulator/superconductors with the ordinary inversion
symmetry~\cite{inversion1,inversion2,inversion3}.

This heuristic picture of the Hopf insulator in Fig.\ref{hopf}
points the possible direction of its experimental realization. The
Hopf insulator can be naively viewed as layers of Chern insulators
stacked along a ring in the momentum space (with a nontrivial
winding along the ring). A $3d$ Weyl
semimetal~\cite{weylsemimetal} can be viewed as layers of Chern
insulators stacked along a line in the momentum space, and the
Chern insulator terminates at the momentum layers with the Weyl
points. Now if we can take a Weyl semimetal with two pairs of Weyl
points in the momentum space, such as the material
MoTe$_2$~\cite{daiweyl}, and annihilate the Weyl points to connect
the ``Chern lines" into a ring, then this system could effectively
become a Hopf insulator, with a proper winding of the Hamiltonian
along the ring. Its boundary Fermi rings are just connected Fermi
arcs of the Weyl semimetal.

Now let us move on to the $4d$ model. A $4d$ band structure is not
just for pure theoretical interest, we can also view the fourth
momentum as the time coordinate, and thus the entire band
structure as a time-dependent $3d$ Hamiltonian. In $4d$, the set
of maps $S^4 \to \Gr(n,m+n)$ is classified by $\pi_4[\Gr(n,m+n)]$.
We will start with the minimal model $m = n = 1$ . In this case,
the band insulator is a map $S^4 \to S^2$, which has homotopy
group $\pi_4[S^2] = Z_2$.

We need to construct a nontrivial map $F\colon T^4 \to S^2$.
Heuristically this mapping can be viewed as the following: in the
$4d$ space, the domain wall between $n_3 > 0$ and $n_3 < 0$ will
form a three-torus $T^3$, and $(n_1, n_2)$ winds nontrivially
along all three directions of the three-torus. Thus the $4d$ Hopf
insulator constructed with the three component vector $\vec{n}$
and Pauli matrices, can be heuristically viewed as
``Chern-insulator$\rtimes T^2$". Consequently, the $3d$ boundary
states will have a torus of zero energy modes (no symmetry is
needed in this minimal two-band model).

\begin{figure}[!thb]
\centering
\includegraphics[width=200pt]{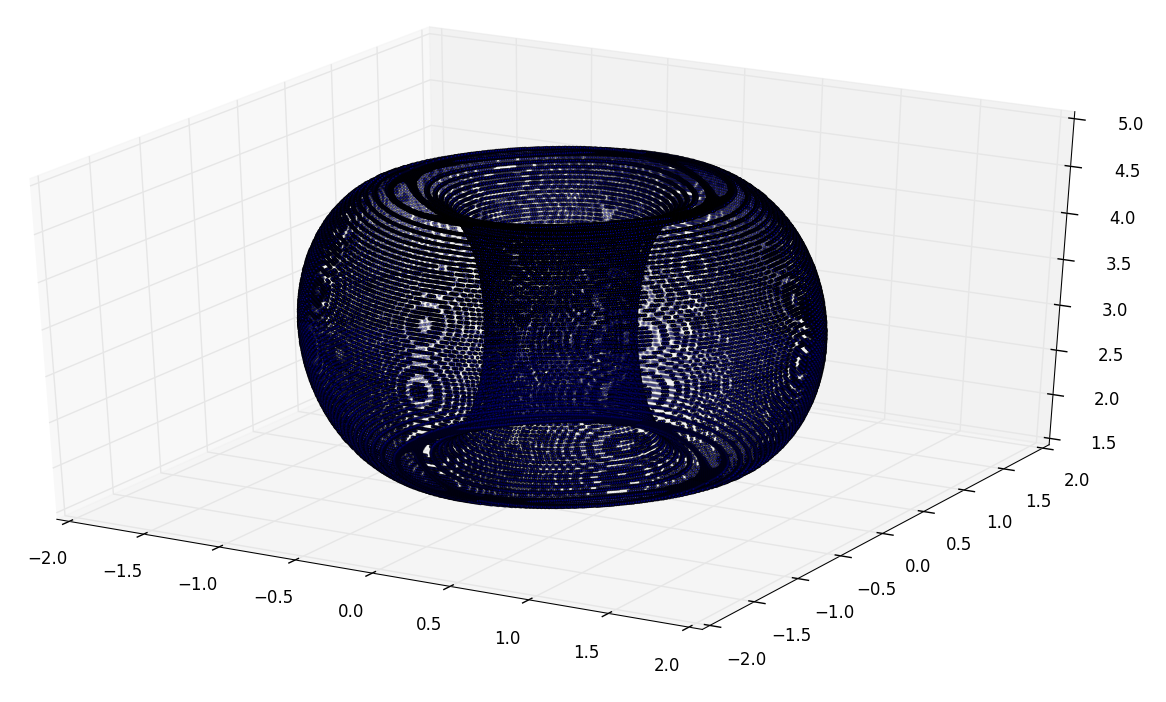}
\caption{The boundary zero energy states of the $4d$ minimal
two-band model for Hopf insulator with $m=2$ in Eq.~\ref{Ns},
plotted in the $3d$ boundary Brillouin zone. Because this model
can be heuristically viewed as ``Chern-insulator$\rtimes T^2$, its
boundary has a torus of zero energy states.}\label{fermisheet}
\end{figure}

A concrete band structure of this kind was discussed in
Ref.~\onlinecite{phdthesis}. We review the idea but implement it
somewhat differently and also generalize it. $F$ can be
constructed via the reduced suspension technique in algebraic
topology \cite{Hatcher,Wilczek}. Pictorially,
$$T^4\xrightarrow{\Sigma [f]} \Sigma S^2=S^3 \xrightarrow{f}S^2$$
where $\Sigma [f]$ ~\cite{T'} is the reduced suspension of the Hopf map
$f$, defined as \beqn \Sigma f\colon (\bold k, t) \mapsto (N_1, \
N_2, \ N_3, \ N_4), \eeqn where $\vec{N}$ is a four component
vector with nonzero norm: \beqn N_1 &=& \sin(t/2)(\sin k_x \sin
k_z \cr &+& \sin k_y (m - \cos k_x - \cos k_y - \cos k_z)), \cr\cr
N_2 &=& \sin(t/2)(\sin k_x (m - \cos k_x  -\cos k_y- \cos k_z) \cr
&-& \sin k_y \sin k_z), \cr\cr N_3 &=& \cos t (\sin^2 k_x + \sin^2
k_y)+ \sin^2 k_z \cr &+& (m - \cos k_x -\cos k_y- \cos k_z)^2,
\cr\cr N_4 &=& \sin t ((m-\cos k_x  -\cos k_y- \cos k_z)^2 \cr &+&
\sin^2 k_z).\label{Ns}\eeqn The Hopf map, as before, is defined as
$f\colon (N_1,N_2,N_3,N_4) \mapsto (n_1,n_2,n_3)$,
\beqn n_1(\bold k, t, m) &=& 2(N_1 N_3 +  N_2  N_4 ) \cr n_2(\bold
k, t, m) &=& 2(N_1 N_4 -  N_2  N_3) \cr n_3(\bold k, t, m) &=&
N_1^2 +  N_2^2 -  N_3^2 - N_4^2. \eeqn Finally, we define the
Hamiltonian $H(\bold k, t, m)$ (up to normalization) as
\beqn H(\bold k, t, m) = \vec n(\bold k, t,m) \cdot \vec \sigma.
\eeqn
The variable $t$ in this $4d$ model can be viewed as time. This
means we can consider our system as an adiabatic time-dependent
band insulator, with a period of $2\pi$. In Fig.~\ref{fermisheet}
we plot the zero energy states computed numerically at the
boundary Brillouin zone of the minimal two-band model of the $4d$
Hopf insulator. The zero energy states indeed form a torus in the
boundary Brillouin zone, which is consistent with our expectation.

We now consider a general multi-band system, and impose the
$\mathcal C^\prime$ symmetry. With the $\mathcal{C}^\prime$
symmetry, there is still an even number of bands, and just like
the $3d$ story discussed before, since $\pi_4[\Sp(N)/\U(N)] = Z_2$
for all $N$, there is only one class of nontrivial Hopf insulator
with the $\mathcal{C}^\prime$ symmetry.

In $4d$ there is a well-known integer quantum Hall state without
assuming any symmetry other than the charge
conservation~\cite{huzhang}, but unlike the $2d$ Chern-insulator,
the $4d$ integer quantum Hall state necessarily breaks the
$\mathcal{C}^\prime$, because its response to the external
electromagnetic field $j^e_\mu \sim
\epsilon_{\mu\nu\rho\tau\sigma} F_{\nu\rho}F_{\tau\sigma}$ breaks
$\mathcal{C}^\prime$, where $j^e_\mu$ is the charge current.

In summary, we found a class of $3d$ and $4d$ topological
insulators whose topological nature is characterized by the Hopf
map and its generalizations. We identified a $\mathcal{C}^\prime$
symmetry which gives these states a $\mathbb{Z}_2$ classification.
The states we constructed are also mathematically equivalent to
topological superconductors with total spin $S_z$ conservation
(but there is no charge conservation). Now the
$\mathcal{C}^\prime$ symmetry becomes a special inversion symmetry
$\mathcal{I}^\prime$ which is a product of the ordinary inversion
and spin $S_z$ flipping. Thus our system can also be viewed as a
crystalline topological superconductor with the $S_z$ conservation
and the $\mathcal{I}^\prime$ symmetry.

We gratefully acknowledge support from the Simons Center for
Geometry and Physics, Stony Brook University, where some of the
research for this paper was performed during the 2016 Simons
Summer Workshop. The authors are supported by the David and Lucile
Packard Foundation and NSF Grant No. DMR-1151208. The authors
thank Sergio Cecotti, Dan Freed, Mike Hopkins, Joel E. Moore, and
Ashvin Vishwanath for very helpful discussions. The first two
authors of this work contributed equally.

\bibliography{hopf}

\begin{thebibliography}{18}
\bibitem[{con()}]{convention}
\bibinfo{note}{In our convention, the rank of Sp($2N)$ is $N$}.
\bibitem[{T'()}]{T'}
\bibinfo{note}{At this point, if we impose $\mathcal T' = \mathcal T \times
  \mathcal I$ symmetry, where $\mathcal T$ is time-reversal such that $\mathcal
  T^2 = -1$, then it can be shown that the configuration space is reduced from
  Sp($4N$)/U($2N$) to Sp$(2N)$, and $\Sigma f$ is nontrivial element of
  $\pi_4[\text{Sp}(2N)] = \mathbb Z_2$ in the case $N = 1$}.
\end{thebibliography}

\newpage

\end{document}